\documentclass[12pt]{article}

\input psfig

\def\xslide#1#2#3#4#5#6#7{\centerline{\psfig
{figure=#1,height=#2,bbllx=#3bp,bblly=#4bp,bburx=#5bp,bbury=#6bp,width=#7,clip=}}}

\title{Three channel model of meson-meson scattering and scalar meson 
       spectroscopy}
\author{R. Kami\'nski$^1$, L. Le\'sniak$^2$\\
Department of Theoretical Physics\\
H. Niewodnicza\'nski Institute of Nuclear Physics,\\ PL 31-342
Krak\'ow, Poland\\and\\
B.\ Loiseau\\ 
Division de Physique Th\'eorique$^*$,
Institut de Physique Nucl\'{e}aire,\\  
F-91406, Orsay Cedex\\ 
and\\ LPTPE Universit\'e P. \& M. Curie, 4 Place Jussieu,\\
             F-75252, Paris Cedex 05, France}

\newcommand {\bfq}{{\bf q}}
\newcommand {\bfp}{{\bf p}}

\newcommand {\bfu}{{\bf u}}
\newcommand {\bfk}{{\bf k}}
\newcommand {\la}{\lambda}
\newcommand {\eq}{\begin{equation}}
\newcommand {\qe}{\end{equation}}
\newcommand{\ba}{\begin{eqnarray}}
\newcommand{\ea}{\end{eqnarray}}
\newcommand{\pp}{$\pi\pi$ }
\newcommand{\kk}{$K\overline{K}$ }
\newcommand{\fo}{$f_0(980)$ }
\newcommand{\epw}{$f_0(1400)$ }
\newcommand{\epsig}{$f_0(500)$ }

\newcommand{\reactpol}{$\pi^- p_{\uparrow} \rightarrow \pi^+ \pi^- n$ }

\newcommand{\roro}{$\sigma\sigma$ }
\newcommand{\fourpi}{$4\pi$ }

\begin{document}

\maketitle


\begin{abstract}
New solutions on the scalar -- isoscalar \pp 
phase shifts are analysed together with previous \kk results using 
a separable potential model of three coupled channels
(\pp, \kk and an effective $2\pi 2\pi$ system).
Model parameters are fitted to two sets of solutions obtained in a recent
analysis of the CERN-Cracow-Munich measurements of the \reactpol
reaction on a polarized target.
A relatively narrow (90 -- 180 MeV) scalar resonance $f_0(1400-1460)$ is found,
 in contrast 
to a much broader ($\Gamma \approx 500$ MeV) state emerging from the analysis of previous 
unpolarized target data.




\end{abstract}


The spectrum of scalar mesons is still not well known 
\cite{pdg96,landua}.
The $J^{PC} = 0^{++}$ meson nonet is not free from ambiguities and
the internal structure of scalar mesons is rather controversial.
Recently, many papers have been devoted to the study of the lowest
scalar glueball 
[3--6].
In these articles properties of the
$f_0(1500)$ resonance, observed recently by the Crystal Barrel Collaboration
in $p\bar p$ annihilations 
\cite{amslerklkl,amsler4pi},
are analysed and interpreted. 
Furthermore, lattice QCD calculations predict the mass of the lowest scalar
glueball to fall between 1500 and 1700 MeV 
\cite{bali,weingarten}. 
One should also
mention increasing evidence for a broad scalar--isoscalar resonance
$\sigma$ 
[11--14]
below 1000 MeV. This state reappeared as $f_0(400 - 1200)$ in the
last edition of the Particle Data Group \cite{pdg96}, 
after 22 years of absence.
Well established resonance \fo lies at the \kk threshold.
Its interpretation, however, is a subject of lively discusions, 
as can be seen for example in Ref. \cite{locher} and articles cited therein.

In Ref. \cite{klm} the scalar meson spectrum was studied in terms of
a relativistic \pp and \kk coupled channel 
model from the $\pi\pi$ threshold up to 1400 MeV. 
The phenomenological parameters were constrained by fitting the
$S$-wave data extracted from the experimental cross sections on the 
$\pi^+\pi^-$ production by $\pi^-$ scattering on unpolarized hydrogen target 
\cite{grayer}.
Further constraints were
imposed by the $K\bar K$ phase shift analysis of Ref. 
\cite{cohen}.

Recently, authors of Ref.
\cite{klr}
have analysed data obtained on
a polarized target by the CERN-Cracow-Munich group for the
$\pi^-p_{\uparrow} \to \pi^+\pi^-n$ reaction 
\cite{becker}. Separation
of the $\pi$ and $a_1$ exchange amplitudes in this reaction was
then possible for the first time, using assumptions much weaker
than in all previous analyses. From a set of four solutions for
the isoscalar $S$-wave phase shifts up to 1600 MeV, two of them
("down-flat" and "up-flat") satisfy the unitarity constraint. The
"down-flat" solution is in good agreement with the former solution
of Ref. \cite{grayer} up to 1400 MeV. Above 1400 MeV one observes an
increase of the phase shifts and larger inelasticies than those
in Ref. \cite{grayer}. 
This could be a manifestation of the presence of 
scalar mesons $f_0(1370)$ or $f_0(1500)$ in that energy range.
There, a strong four--pion production has been
observed in different experiments 
[7,8,20--22].
This provides a compelling argument to take into
account the $4\pi$ channel. In this channel there is some
evidence of clustering into $\sigma\sigma$ or $\rho\rho$ pairs
[6--9].
In \cite{kloet} an attempt was made to describe the $\pi\pi$ $S$-wave in 
the 1000 to 2000 
MeV region by generalization of the non-relativistic 2--channel model 
\cite{cdl88} to three coupled channels. 

 In the present paper we extend the isospin 0 $S$--wave relativistic 2--channel
model of Ref. \cite{klm} by adding to its $\pi\pi$ and $K\bar K$
channels an effective third coupled channel, here
called $\sigma\sigma$.


We consider the $S$-wave scattering and transition reactions between three 
coupled channels of meson pairs labelled 1, 2 and 3. Reaction amplitudes 
$T$ satisfy a system of coupled channel Lippmann-Schwinger equations
\cite{klm}:
\eq 
<\bfp|T|\bfq>=<\bfp|V|\bfq>+ \int \frac{d\bfu}{(2\pi)^3} <\bfp|V|\bfu>
                  <\bfu|G|\bfu><\bfu|T|\bfq>, 
\label{tpq}\qe
where $V$, $G$ and $T$ are $3 \times 3$ matrices, $V$ is the interaction matrix 
and $G$ is the diagonal matrix of channel propagators:
\eq 
G_j(E,\bfu)=\frac{1}{E-2E_j(\bfu)+i\epsilon},\ \ \  \epsilon \to 0(+), \ \ \ j=1,2,3. 
\label{pro}\qe
In  Eq. (\ref{pro}) $E$ is the total energy, $E_j= \sqrt{m_j^2+\bfu^2}$ and $m_j$ 
is the meson mass in channel $j$.  We consider meson pairs of same 
mass and momentum {\bf u} in their centre of mass system.

We choose a separable form of the interaction:
\eq <\bfp|V_{\alpha \gamma,\,j}|\bfq> = \sum_{j=1}^{n}\la_{\alpha \gamma,\,j}\ 
                             g_{\alpha,\,j}(\bfp)\ g_{\gamma,\,j}(\bfq),                  
                              \ \ \alpha, \gamma = 1, 2, 3,
                              \label{pot}\qe
where $\la_{\alpha \gamma,\,j}$ are coupling constants and 
\eq  g_{\alpha,\,j}(\bfp)= \sqrt{\frac{4\pi}{m_j}}
                             \frac{1}{\bfp^2+(\beta_{\alpha,\,j})^2}
                             \label{for}\qe
are form factors which depend on the relative centre of mass meson 
momenta $\bfp$ in the final channel or $\bfq$ in the initial channel.  In the
\pp channel ($j=1$) we choose a rank-2 separable potential ($n=2$) and in
the other channels, i.e.  $K \overline{K} \ (j= 2)$ and $\sigma \sigma \ (j=3)$, 
a rank-1 potential ($n=1$).

Following the formalism developed in Ref. \cite{klm} we can solve the system of 
equations (\ref{tpq}) -- (\ref{for}), which leads to the Jost function
\eq D(\bfk_1, \bfk_2, \bfk_3) = det(1-\la I), \label{jos}\qe
where $\la$ and $I$ are symmetric $4 \times 4$ matrices

\eq \la = \left(
\begin{array}{cccc}
 \la_{11,1} &             0  & \la_{12,1} & \la_{13,1} \\
            0   & \la_{11,2} & \la_{12,2} & \la_{13,2} \\
 \la_{12,1} & \la_{12,2} & \la_{22,1} & \la_{23,1} \\
 \la_{13,1} & \la_{13,2} & \la_{23,1} & \la_{33,1} \\
\end{array}\right),\ 
 I = \left(
\begin{array}{cccc}
I_{1,11} & I_{1,12} &    0         &       0      \\
I_{1,12} & I_{1,22} &    0         &       0      \\
    0        &     0        & I_{2,11} &       0      \\
    0        &     0        &    0         & I_{3,11} \\
\end{array}\right) \qe

and

\eq I_{\alpha,ij}(\bfk_\alpha) = \int \frac{d\bfu}{(2\pi)^3}\ 
                                   g_{\alpha,i}(\bfu)\ G_\alpha(E,\bfu)\ 
                                   g_{\alpha,\,j}(\bfu).
                                   \label{ial}\qe
In (\ref{jos}) and (\ref{ial}) $\bfk_\alpha$ are the on-shell momenta connected 
with the total energy by
\eq E = 2\sqrt{\bfk_1^2+m_1^2} = 2\sqrt{\bfk_2^2+m_2^2} = 2\sqrt{\bfk_3^2+m_3^2}.
                              \label{eto}\qe
Analytical expressions for $I_{\alpha,ij}(\bfk_\alpha)$ integrals are
given in Appendix  A of Ref. \cite{klm}.  Altogether, this model has 13
parameters:  9 coupling constants $\la_{\alpha \gamma,\,j}$ and 4 range
parameters $\beta_{\alpha,\,j}$ (for $\alpha = 1\ j$ = 1 or 2 and for $\alpha
\neq 1 \  j=1$). In the calculations we shall use dimensionless coupling
constants defined as 
\eq \Lambda_{\alpha\gamma,\,j} =
\frac{\lambda_{\alpha\gamma,\,j}}{2\left(\beta_{\alpha,\,j} \beta_{\gamma,\,j}\right)^{3/2}}.\qe
From now on, we will omit index $j$ for $\lambda_{\alpha \gamma, \,j}$ 
if both $\alpha$ and $\gamma$ are different from 1 and for 
$\beta_{\alpha,\,j}$ if $\alpha \neq 1$.

$S$-matrix elements $S_{\alpha \beta}\ (\alpha, \beta = 1, 2, 3)$ can be 
written in terms of the Jost function of different arguments, for example
\eq S_{11} = \frac{D(-\bfk_1, \bfk_2, \bfk_3)}{D(\bfk_1, \bfk_2, \bfk_3)}.
              \label{s11}\qe                     
Further expressions for some other matrix elements can be found in Ref. 
\cite{lles96}.
The model satisfies the unitarity condition $S^+S = 1$.
Diagonal matrix elements are parametrized as
\eq S_{jj} = \eta_je^{2i\delta_j}, ~~j = 1,2,3,\qe
where $\eta_j$ and $\delta_j$ are channel $j$ inelasticities and 
phase shifts, respectively. Expressions for nondiagonal elements can be
found in \cite{lles96}. 
Some of the $S$-matrix poles in the
complex energy plane can be interpreted as resonances. 
They correspond to the zeroes of the Jost function
$D(\bfk_1,\bfk_2,\bfk_3)$ as can be seen from (\ref{s11}). 


We fit the existing experimental results on the
$\pi\pi$ $S$-wave isoscalar phase shifts together with inelasticity in
the $\pi\pi$ channel and with the $K\bar K$ phase shifts. In this
analysis we extend the 2--channel model developed in Ref. \cite{klm} since
new data have been recently obtained \cite{klr} in a wider mass range
(up to 1600 MeV).
We choose a method based on the $\chi^2$ fit:
\eq \chi^2 = \chi^2_{\pi} + \chi^2_{\pi K} + \chi^2_{\eta},\qe
\eq \chi^2_{\pi} = \sum_{l=1}^{N_\pi}
\left\vert\frac{e^{2i\delta^{th}_{\pi\pi}} -
e^{2i\delta^{exp}_{\pi\pi}}}{2\Delta\delta_{\pi}}\right\vert^2, \label{chipi}\qe
\eq \chi^2_{\pi K} =  \sum_{l=1}^{N_{\pi K}} 
\left\vert\frac{e^{2i(\delta^{th}_{\pi\pi}+\delta^{th}_{KK})}
-e^{2i(\delta^{exp}_{\pi\pi}+\delta^{exp}_{KK})}}{2\Delta(\delta_{\pi\pi}
+ \delta_{KK})}\right\vert^2,\label{chik}\qe
\eq \chi^2_{\eta} =  \sum_{l=1}^{N_{\eta}}\left(\frac{\eta_{\pi\pi}^{th} -
\eta^{exp}_{\pi\pi}}{\Delta\eta_{\pi\pi}}\right)^2,\label{chieta}\qe
where $\delta_{\pi\pi}, \delta_{KK}$ are phase shifts in
channels 1 and 2, $\Delta \sigma_{\pi}, \Delta \sigma_{K}$ are
the corresponding experimental errors and $\eta_{\pi\pi},
\Delta\eta_{\pi\pi}$ are inelasticities and inelasticity errors,
respectively. Superscripts "th" or "exp" refer to
theoretical or experimental values. In Eqs. (\ref{chipi}) to (\ref{chieta})
 $N_{\pi}, N_{\pi K}$ and $N_{\eta}$ are the numbers of experimental points.

Below 600 MeV we use data from the $K_{e4}$ decay \cite{rosselet} and from
Refs. \cite{belkov} and \cite{srinivasan}. Above 600 MeV we use the "down-flat"
and "up-flat" solutions of the analysis of Ref. \cite{klr} on a
polarized target. Constraints in
the $K^+K^-$ channel are also needed. 
  Therefore, we have used the results of the analysis of reactions
$\pi^-p \to K^+K^-n$ and 
$\pi^+n \to K^-K^+p$  \cite{cohen}, although
targets were unpolarized there.  


We begin with a simpler $\chi^2$-minimization by
considering only the $\pi\pi$ and $K\bar K$ 2--channel case.
Starting parameters of the interaction potentials have been taken from 
2--channel fits obtained in \cite{PhD}.
In this case there are 8 free  parameters, 
as indicated in Table \ref{parameters}. 
The corresponding fits for $\delta_{\pi\pi},
\eta_{\pi\pi}$ and $\varphi_{\pi K}=
\delta_{\pi\pi}+\delta_{KK}$, plotted as dotted lines, are
compared to the experiment in Figs. 1, 2 and 3.
Above 1400 MeV both "down-flat" and
"up-flat" data indicate a decrease towards small values of $\eta\ 
(\eta \approx 0.6$ to $0.7)$, albeit with large errors.
Furthermore, it can be seen in Fig. 2 that
the errors of $\eta$ are much larger than dispersion of their values. 
Therefore, in order to reproduce more easily this trend of $\eta$ as a function 
of energy, for most of the data points we reduce in Eq. (15) the experimental 
$\eta$ errors to 0.1.
For the "down--flat" solution we keep, however, the original errors at 1310, 
1470 and 1570 MeV since inelasticities at these energies are
above or close to 1.    
For the last point at 1590 MeV we increase the error from 0.41 to 1 since 
$\eta$ is as large as 1.52. Similarly, for the "up-flat" solution we do not
change the errors in the range between 1350 and 1450 MeV, however, at 1310, 1470
and 1590 MeV we increase the errors to 0.8 since $\eta$ values at those energies
are substantially higher than 1.

We found that in the 2--channel case for the "down-flat" solution it was not
possible to obtain the $\eta$ values with such modified errors
between 1400 to 1600 MeV
(this is slightly different for the "up--flat" solution as can be seen in Fig. 2b).
In the 3--channel model, however, we can get a substantial
decrease of $\eta$ above 1400 MeV (see Fig. 2a). 
In order to achieve this behaviour, couplings between the $\pi\pi$ and
$\sigma\sigma$ or \kk channels should be sizable. 
The corresponding 14 free parameters including the $m_3$ mass for two of our best 
fits are given in Table
\ref{parameters}, and their $\chi^2$ in Table \ref{chi_sqrd}. 
The 3--channel
fits for the "down-flat" data are labelled by A and B, and those
corresponding to the "up-flat" data - by C and D, respectively.
The fits favour $m_3$ masses in the range of 675 to 733 MeV.
The values of $\delta_{\pi\pi}, \eta_{\pi\pi}$ and $\varphi_{\pi K}$ are drawn as
thick and thin solid lines in Figs. 1-3 and compared to data.
In Fig. 1 we have shown fits A and C only since energy dependence
of fit B is very close to that of  A, and that of D -- very close to C. 
In Fig. 1 one can already notice 
a better agreement with data of the 3--channel model in comparison with the 
2--channel one. This is especially well visible in Fig. 1b. 
The main difference between the 2-- and 3--channel fits lies in $\eta$
above 1400 MeV, where the opening of the 
$\sigma\sigma$ channel
leads to a fast decrease of inelasticity parameters (see Fig. 2a). Let us also note
an improvement in $\varphi_{\pi K}$ over the 1000 to 1200 MeV range, as can be seen in
Fig. 3 and in Table \ref{chi_sqrd} with a better $\chi^2_{\pi K}$.

Fits of similarly good quality were obtained with very 
different physical parameters in the \kk and \roro channels.
For example, in the 2--channel fits and 
in fit A, the \kk interaction is attractive while 
in the other cases it is repulsive (Table \ref{parameters}). 
Similarly, interchannel couplings are very different in both cases.
In fit A we see rather strong, and in fit C  
very strong \pp to \roro  and \kk to \roro
couplings, while in the B and D cases the \pp -- \kk coupling, $\Lambda_{12,2}$,
is particularly strong.
Lack of a sufficient number of observables and/or experimental precision, in
particular in the effective $4\pi$ channel, leads to the existence of several good
sets of model parameters with quite different channel and interchannel 
interactions. Other 3--channel fits of reasonable quality with less than
14 parameters can be also obtained.

We have studied positions of the $S$--matrix poles in the complex energy plane
($E = M- i\Gamma/2$).
For the 3--channel model there are 8  different sheets 
which correspond to different signs of imaginary 
parts of channel momenta ($Im p_1, Im p_2, Im p_3$). 
For example, on the sheet denoted by ($---$) all imaginary parts are negative.
In the case of the 2--channel model there are only 4 sheets labelled by signs of
$Im p_1$ and  $Im p_2$.
Resonance parameters predicted by the 2-- and 3--channel models are summarized in 
Table 3.
 At low energy we find a very broad \epsig resonance 
(also called $\sigma$ meson).
Since the "up--flat" data indicate a faster increase of the \pp phase shifts
with energy than the
"down--flat" data, the $\sigma$ meson appears in the first case at a mass
higher
by about 40 MeV and with a width smaller by about 50 MeV. 
In the three channel fits the \fo resonance is seen in the vicinity of
the \kk threshold with a width of about 60 to 70 MeV.
The relatively narrow state \epw appears in the 3--channel fits (see 
Table \ref{resonances}). Its mass varies from about 1400 MeV to 1460 MeV.
For the "down--flat" fits this resonance is narrower 
($\Gamma \approx 100$ MeV) on sheet $(--+)$ than on sheet $(---)$.
This width is very close to values found by the 
Crystal Barrel Group (Refs. \cite{amslerklkl,amsler4pi}) but the resonance 
masses are smaller than their values close to about 1500 MeV.
We should point out that a narrow ($\Gamma= 65 \pm 10$ MeV) scalar resonance at
$M=1445 \pm 5$ MeV has been found by the WA91 group at CERN in central 
production of \fourpi in high--energy $pp$ collisions \cite{antinori}.

Finally, let us mention an important  difference between the fits
presented here for data 
taken on a polarized target in comparison with those performed with data 
\cite{grayer}  
obtained on a nonpolarized target.
If we fit the data of \cite{grayer} using our 3--channel model, we obtain a
very wide resonance at $M=1521$ MeV of width 503 MeV. This means that the recent 
analysis of data \cite{klr} supplies some new information on the \epw meson.

In conclusion, we have analysed the isoscalar S--wave \pp and \kk scattering
using the CERN--Cracow--Munich data \cite{klr} together 
with the data of 
[17,26--28]
in the framework of
the 2- and the 3--channel models of meson--meson scattering. 
All fits of the phase shifts analysis \cite{klr} indicate
presence of a relatively narrow (90 -- 180 MeV) scalar resonance of mass
1400 -- 1460 MeV. This resonance is quite compatible with recent 
observations of a possible scalar glueball at 1500 MeV with a width of 100 MeV.

\vspace{0.5cm}
We are indebted to D. V. Bugg for useful correspondence
and to J. Kisiel, V. E. Markushin and K. Rybicki for fruitful discussions.

This work has been performed in the framework of the IN2P3 -- Polish Laboratories
Convention (project No 93-71) and partially supported by the Polish State 
Committee for Research (grants No 2 P03B 231 and No 2 P03B 020 12). 

\vspace{0.5cm}

-----------------------------

$^1$ E-mail: kaminski@solaris.ifj.edu.pl

$^2$ E-mail: lesniak@bron.ifj.edu.pl

$^*$ Unit\'e de 
Recherche des Universit\'es Paris 11 et Paris 6 Associ\'ee au CNRS 


\begin{table}[h]
\centering
\caption{Separable interaction parameters for 2-- and 3--channel model fits
to the "down-flat" and "up-flat" data from [18]. 
Values of $\beta$ and $m_3$ are given in GeV.} 

\vspace{.7cm}

\begin{tabular}{|l|c|c|c|}
\hline
\multicolumn{1}{|c|}{Data} &
\multicolumn{3}{|c|}{down--flat}   \\    
\hline 
\multicolumn{1}{|c|}{model} &
\multicolumn{1}{|c|}{2-channel}       &   
\multicolumn{2}{|c|}{3--channel} \\
\hline
\multicolumn{1}{|c|}{fit}        &
\multicolumn{1}{|c|}{}        &   
\multicolumn{1}{|c|}{A}       &
\multicolumn{1}{|c|}{B}       \\   
\hline
$\Lambda_{11,1}$ & $-.14258\times 10^{-3}$ & $-.29975\times 10^{-2}$ & $-.52138\times 10^{-2}$  \\
$\Lambda_{11,2}$ & $-.18895$         & $-.10844$         & $-.10552$          \\
$\Lambda_{22}$   & $-.49106$         & $-.39304$         & $3.1637$           \\
$\Lambda_{33}$   & 0                 & $-.17447\times 10^{-2}$ & $-.45719\times 10^{-1}$  \\
$\Lambda_{12,1}$ & $.27736\times 10^{-5}$  & $.12039\times 10^{-2}$  & $.10685\times 10^{-1}$   \\
$\Lambda_{12,2}$ & $.43637\times 10^{-1}$  & $-.11333$         & $-.80626$          \\
$\Lambda_{13,1}$ & 0                 & $-.25841\times 10^{-2}$ & $.14544\times 10^{-4}$   \\
$\Lambda_{13,2}$ & 0                 & $.39924$          & $.21878\times 10^{-2}$   \\
$\Lambda_{23}$   & 0                 & $-.57955$         & $-.17515\times 10^{-1}$  \\
$\beta_{1,1}$    & $3.0233\times 10^3$     & $1.426\times 10^{2}$    & $.81615\times 10^2$      \\
$\beta_{1,2}$    & $1.0922$          & $.92335$          & $.85776$           \\
$\beta_2$        & 2.2941            & $1.4959$          & $.47403$           \\
$\beta_3$        & --------          & $1.3676$          & $.45357\times 10^2$      \\
$m_3$            & --------          & .70               & .67510             \\
\hline
\hline\hline
\hline
\multicolumn{1}{|c|}{Data} &
\multicolumn{3}{|c|}{up--flat}   \\    
\hline 
\multicolumn{1}{|c|}{model} &
\multicolumn{1}{|c|}{2-channel}       &   
\multicolumn{2}{|c|}{3--channel} \\
\hline
\multicolumn{1}{|c|}{fit}        &
\multicolumn{1}{|c|}{}        &   
\multicolumn{1}{|c|}{C}       &
\multicolumn{1}{|c|}{D}       \\   
\hline
$\Lambda_{11,1}$ & $-.25550\times 10^{-4}$  & $-.24649\times 10^{-2}$ & $-.13063\times 10^{-2}$  \\
$\Lambda_{11,2}$ & $-.15677$          & $-.73339\times 10^{-1}$ & $-.94483\times 10^{-1}$  \\
$\Lambda_{22}$   & $-.66905$          & .19475            & 3.6391             \\
$\Lambda_{33}$   & 0                  & $-.45073\times 10^{-2}$ & -1.4588            \\
$\Lambda_{12,1}$ & $-.23243\times 10^{-5}$  & $.12710\times 10^{-2}$  & $.10185\times 10^{-2}$   \\
$\Lambda_{12,2}$ & $.74350\times 10^{-1}$   & $-.32171$         &  -.85407           \\
$\Lambda_{13,1}$ &  0                 & $-.52567\times 10^{-2}$ & $-.18043\times 10^{-3}$  \\
$\Lambda_{13,2}$ &  0                 & 1.6134            &   .17803           \\
$\Lambda_{23}$   &  0                 & $-4.0462$         & -.50484            \\
$\beta_{1,1}$    &  $1.6878\times 10^{4}$   & $1.7379\times 10^{2}$   & $3.2899\times 10^2$      \\
$\beta_{1,2}$    &  1.5882            & .92153            &  .90097            \\
$\beta_2$        &  1.4710            & .69032            &  .30733            \\
$\beta_3$        & --------           & .23546            &  1.1419            \\
$m_3$            & --------           & .73252            &  .68096            \\
\hline
\end{tabular}
\label{parameters}
\end{table}


\begin{table}[h]
\centering
\caption{Best $\chi^2$ values of 2-- and 3--channel model fits for the 
"down--flat"
and "up-flat" data of ref. [18]. 
Numbers of data points are indicated in parentheses.
$\overline{\chi}^2$ values are obtained when fitting 
with reduced $\eta$ errors.}

\vspace{.7cm}

\begin{tabular}{|l|c|c|c|c|c|c|}
\hline
\multicolumn{1}{|c|}{Data}       &
\multicolumn{3}{|c|}{down--flat} &
\multicolumn{3}{|c|}{up--flat} \\
\hline
\multicolumn{1}{|c|}{Model}      &
\multicolumn{1}{|c|}{2--channel} &
\multicolumn{2}{|c|}{3--channel} &
\multicolumn{1}{|c|}{2--channel} &
\multicolumn{2}{|c|}{3--channel} \\
\hline
\multicolumn{1}{|c|}{fit}    &
\multicolumn{1}{|c|}{}            &  
\multicolumn{1}{|c|}{A}           &  
\multicolumn{1}{|c|}{B}           &  
\multicolumn{1}{|c|}{}            &  
\multicolumn{1}{|c|}{C}           &  
\multicolumn{1}{|c|}{D}           \\  
\hline 
\multicolumn{1}{|c|}{$\chi_{\pi}^2$ (65)}   & 66.4  & 63.0 & 61.2 & 115.6 & 88.0  & 85.3  \\
\multicolumn{1}{|c|}{$\chi^2_{\pi+K}$ (21)} & 26.3  & 15.9 & 9.7  & 27.0  & 13.1  & 13.2  \\
\multicolumn{1}{|c|}{$\chi_{\eta}^2$ (30)}  & 9.4   & 13.2 & 12.9 & 10.9  & 14.1  & 14.6  \\
\multicolumn{1}{|c|}{$\chi^2_{tot}$ (116)}  & 102.1 & 92.1 & 83.8 & 153.5 & 115.2 & 113.1 \\ 
\hline 
\multicolumn{1}{|c|}{$\overline{\chi_{\eta}}^2$ (30)}  & 100.6 & 36.7  & 29.3  & 43.3  & 36.2  & 37.7   \\
\multicolumn{1}{|c|}{$\overline{\chi_{tot}}^2$ (116)}  & 193.2 & 115.6 & 100.1 & 186.0 & 137.3 & 136.2  \\ 
\hline 
\end{tabular}
\label{chi_sqrd}
\end{table}


\begin{table}[h]
\centering
\caption{Masses and widths of resonances found for the 2-- and 
3--channel fits to the "down--flat" and "up-flat" data}

\vspace{.7cm}

\begin{tabular}{|l|c|c|c|c|c|}
\hline
\multicolumn{6}{|c|}{{\bf 2-channel model}}            \\ 
\hline
\multicolumn{1}{|c|}{pole}         &
\multicolumn{1}{|c|}{sheet}        &
\multicolumn{2}{|c|}{down-flat}    &
\multicolumn{2}{|c|}{up-flat}     \\
\cline{3-6}
\multicolumn{1}{|c|}{}                &
\multicolumn{1}{|c|}{}                &
\multicolumn{1}{|c|}{$M$ (MeV)}       &
\multicolumn{1}{|c|}{$\Gamma$ (MeV)}  & 
\multicolumn{1}{|c|}{$M$ (MeV)}       &
\multicolumn{1}{|c|}{$\Gamma$ (MeV)}  \\ 
\hline
\multicolumn{1}{|c|}{$f_0(500)$ ($\sigma$)}    & $-+$ & 524.0  & 513.6  & 592.9  & 401.8  \\  
\multicolumn{1}{|c|}{$f_0(980)$}               & $-+$ & 993.4  & 79.3  &  1015.0 &  101.4 \\  
\multicolumn{1}{|c|}{$f_0(1400)$}              & $--$ & 1434.6  & 167.6  & 1429.7  & 179.2  \\  
\hline\hline
\hline\hline
\multicolumn{6}{|c|}{{\bf 3-channel model}}            \\
\hline 
\multicolumn{1}{|c|}{}             &
\multicolumn{1}{|c|}{}             &
\multicolumn{4}{|c|}{down-flat}    \\
\cline{3-6}
\multicolumn{1}{|c|}{pole}         &
\multicolumn{1}{|c|}{sheet}        &
\multicolumn{2}{|c|}{A}            &
\multicolumn{2}{|c|}{B}            \\ 
\cline{3-6}
\multicolumn{1}{|c|}{}                &
\multicolumn{1}{|c|}{}                &
\multicolumn{1}{|c|}{$M$ (MeV)}       &
\multicolumn{1}{|c|}{$\Gamma$ (MeV)}  & 
\multicolumn{1}{|c|}{$M$ (MeV)}       &
\multicolumn{1}{|c|}{$\Gamma$ (MeV)}  \\ 
\hline
\multicolumn{1}{|c|}{$f_0(500)$ ($\sigma$)}    & $-++$ & 518.1  & 521.4 & 511.8  & 532.6 \\  
\multicolumn{1}{|c|}{$f_0(980)$}               & $-++$ & 989.0  & 62.0  & 992.4  & 68.2  \\  
\multicolumn{1}{|c|}{$f_0(1400)$:}             & $---$ & 1405.1 & 147.8 & 1411.5 & 169.3 \\  
\multicolumn{1}{|c|}{}                         & $--+$ & 1456.4 & 93.3  & 1402.7 & 108.2 \\  
\hline
\hline\hline
\multicolumn{1}{|c|}{}             &
\multicolumn{1}{|c|}{}             &
\multicolumn{4}{|c|}{up-flat}    \\
\cline{3-6}
\multicolumn{1}{|c|}{pole}         &
\multicolumn{1}{|c|}{sheet}        &
\multicolumn{2}{|c|}{C}            &
\multicolumn{2}{|c|}{D}            \\ 
\cline{3-6}
\multicolumn{1}{|c|}{}                &
\multicolumn{1}{|c|}{}                &
\multicolumn{1}{|c|}{$M$ (MeV)}       &
\multicolumn{1}{|c|}{$\Gamma$ (MeV)}  & 
\multicolumn{1}{|c|}{$M$ (MeV)}       &
\multicolumn{1}{|c|}{$\Gamma$ (MeV)}  \\ 
\hline
\multicolumn{1}{|c|}{$f_0(500)$ ($\sigma$)}    & $-++$ & 561.7  & 466.8 & 558.2  &  477.6 \\  
\multicolumn{1}{|c|}{$f_0(980)$}               & $-++$ & 992.2  & 67.6  & 995.3  &  71.7  \\  
\multicolumn{1}{|c|}{$f_0(1400)$:}             & $---$ & 1407.0 & 181.2 & 1418.8 &  179.3 \\  
\multicolumn{1}{|c|}{}                         & $--+$ & 1423.8 & 177.4 & 1416.5 &  173.2 \\  
\hline
\end{tabular}
\label{resonances}
\end{table}
\begin{figure}[ptb]
\caption{
Energy dependence of $\pi\pi$ phase shifts: 
a) fit to "down-flat" data of [18],
thick solid line corresponds to fit A of the 3--channel model and
dotted line to the 2--channel model fit; 
b) fit to "up-flat" data of [18], 
thick solid line corresponds to fit C of the 3--channel model and
dotted line to the 2--channel model fit.
} 

\vspace{0.5cm}

                \label{pipifig} 
    \begin{center}
\xslide{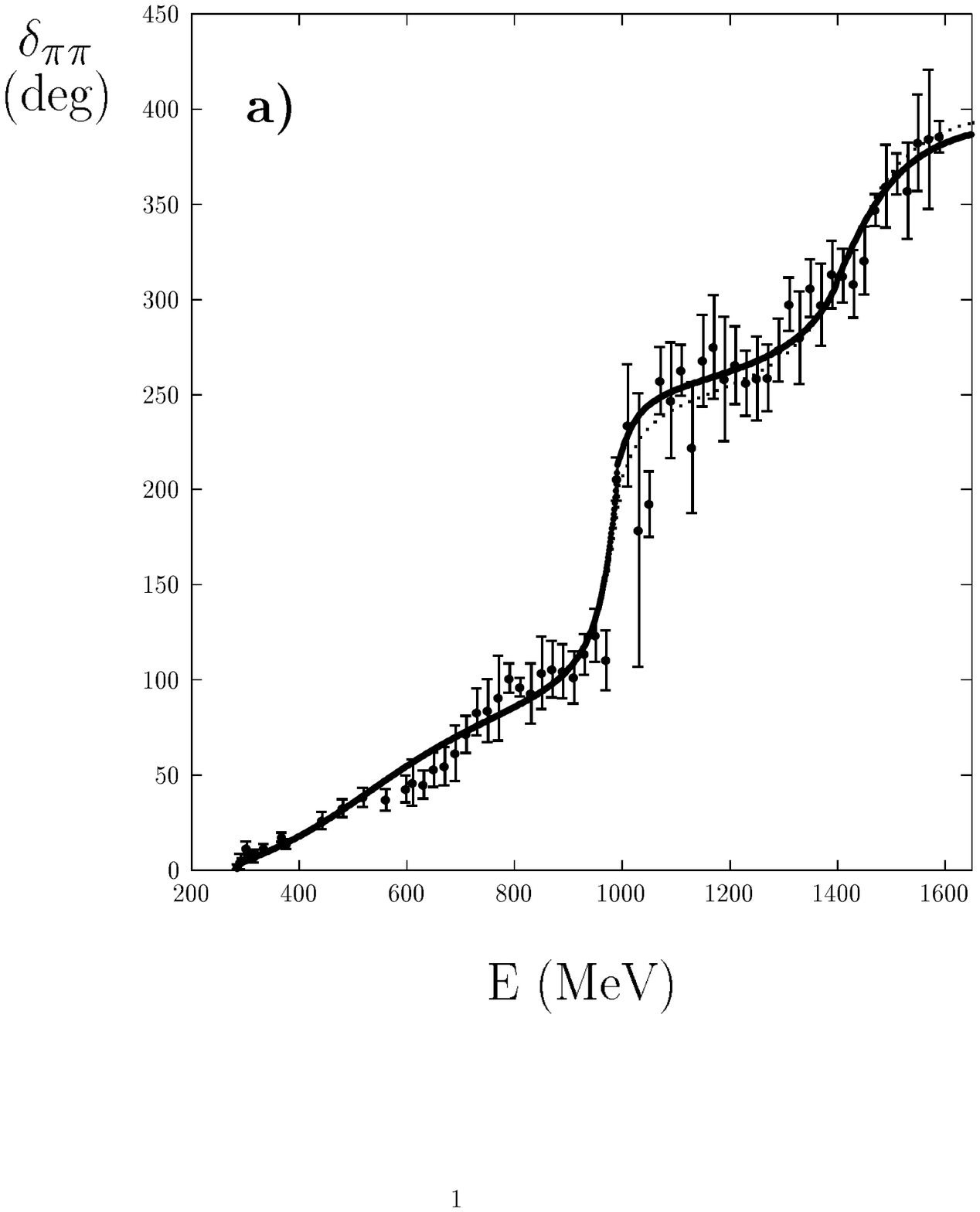}{10cm}{90}{220}{550}{710}{14cm} 
\xslide{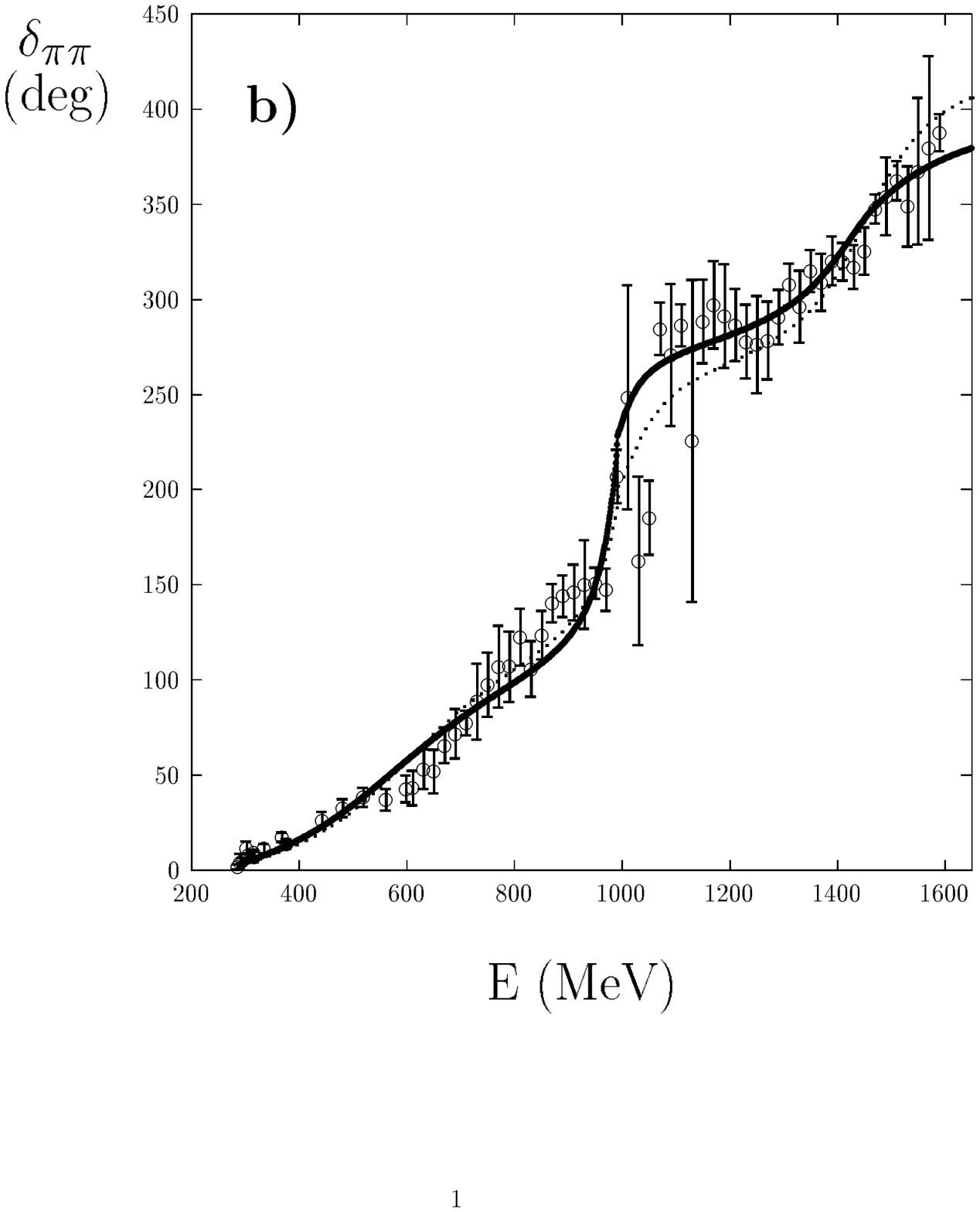}{10cm}{90}{220}{550}{710}{14cm} 
    \end{center}
  \end{figure}
  \begin{figure}[ptb]
\caption{
Energy dependence of inelasticity parameter $\eta_{\pi\pi}$: 
a) fit to  "down-flat" data of [18],
thick solid line corresponds to fit A, thin solid line to fit B and
dotted line to the 2--channel model fit;
b) fit to "up-flat" data of [18], 
thick solid line corresponds to fit C, thin solid line to fit D and
dotted line to the 2--channel model fit.
} 

\vspace{0.5cm}

                \label{etafig} 
    \begin{center}
\xslide{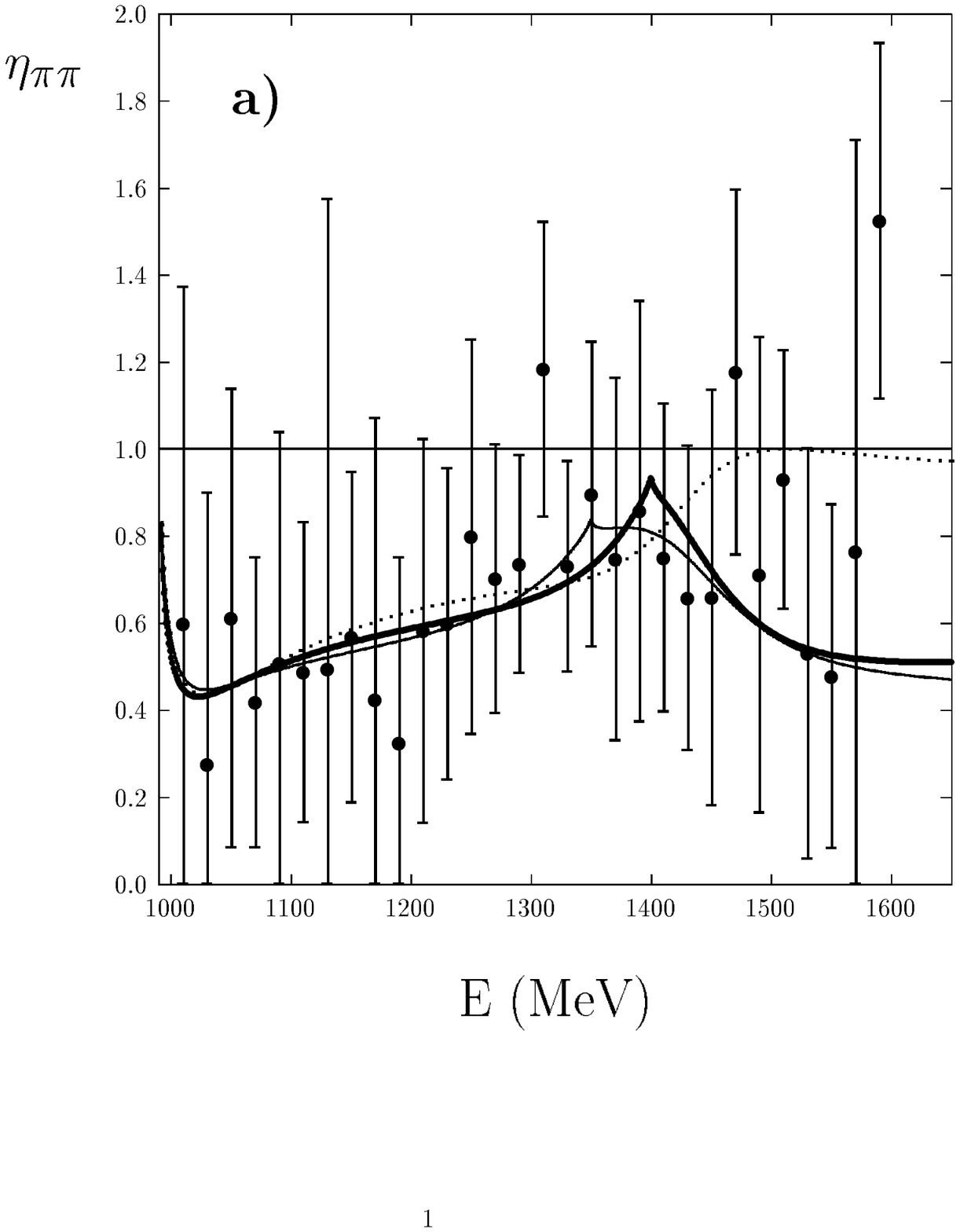}{10cm}{90}{220}{550}{710}{14cm} 
\xslide{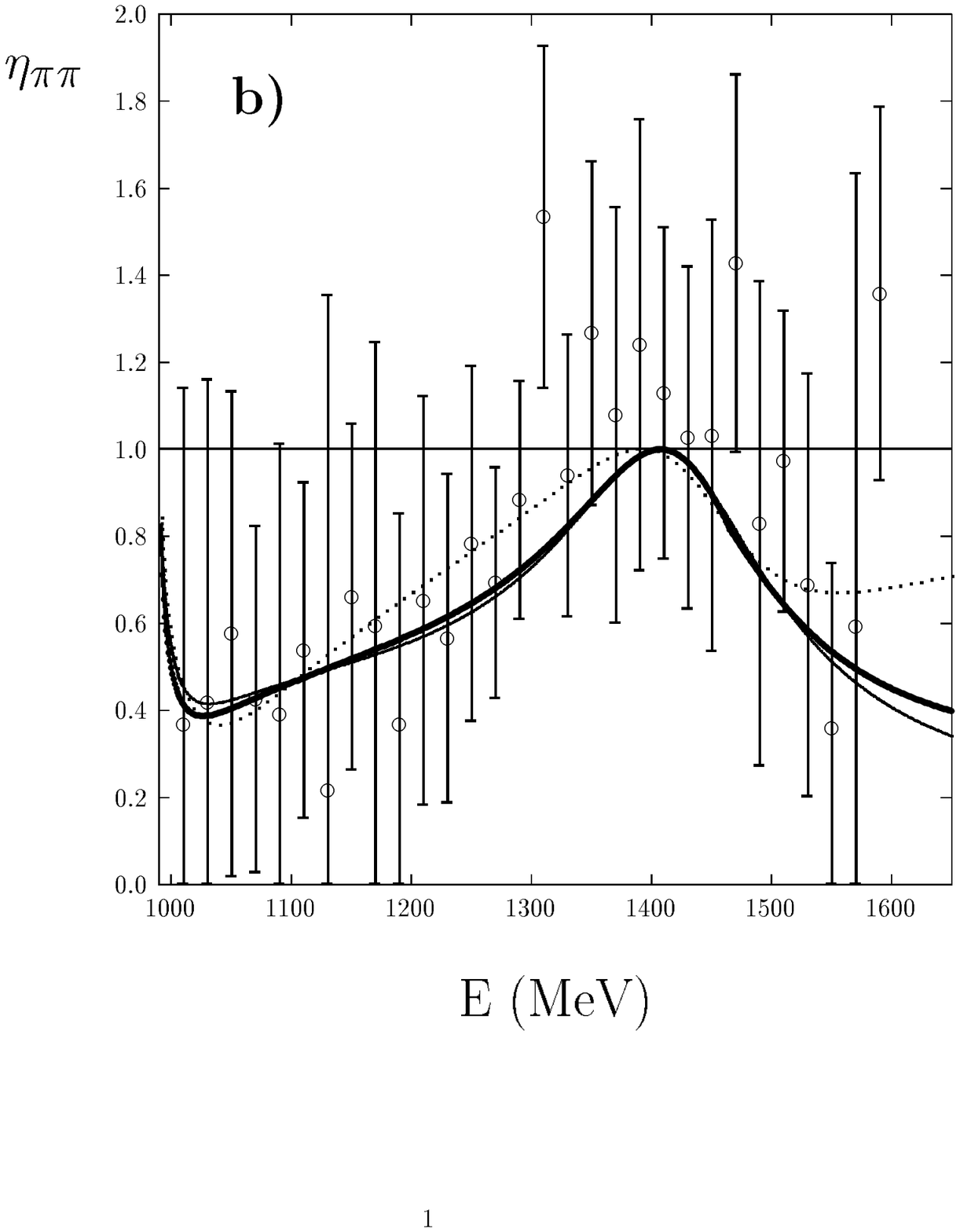}{10cm}{90}{220}{550}{710}{14cm} 
    \end{center}
  \end{figure}
  \begin{figure}[ptb]
\caption{
Energy dependence of phase shifts sum
$\varphi_{\pi K} = \delta_{\pi\pi} + \delta_{K\overline{K}}$:
a) fit to  "down-flat" data of [18],
thick solid line corresponds to fit A, thin solid line to fit B
and dotted line to the 2--channel model fit;
b) fit to "up-flat" data,
thick solid line corresponds to fit C, thin solid line to fit D
and dotted line to the 2--channel model fit.
Data for $\varphi_{\pi K}$ are taken from [17].
} 

\vspace{0.5cm}

                \label{phifig} 
    \begin{center}
\xslide{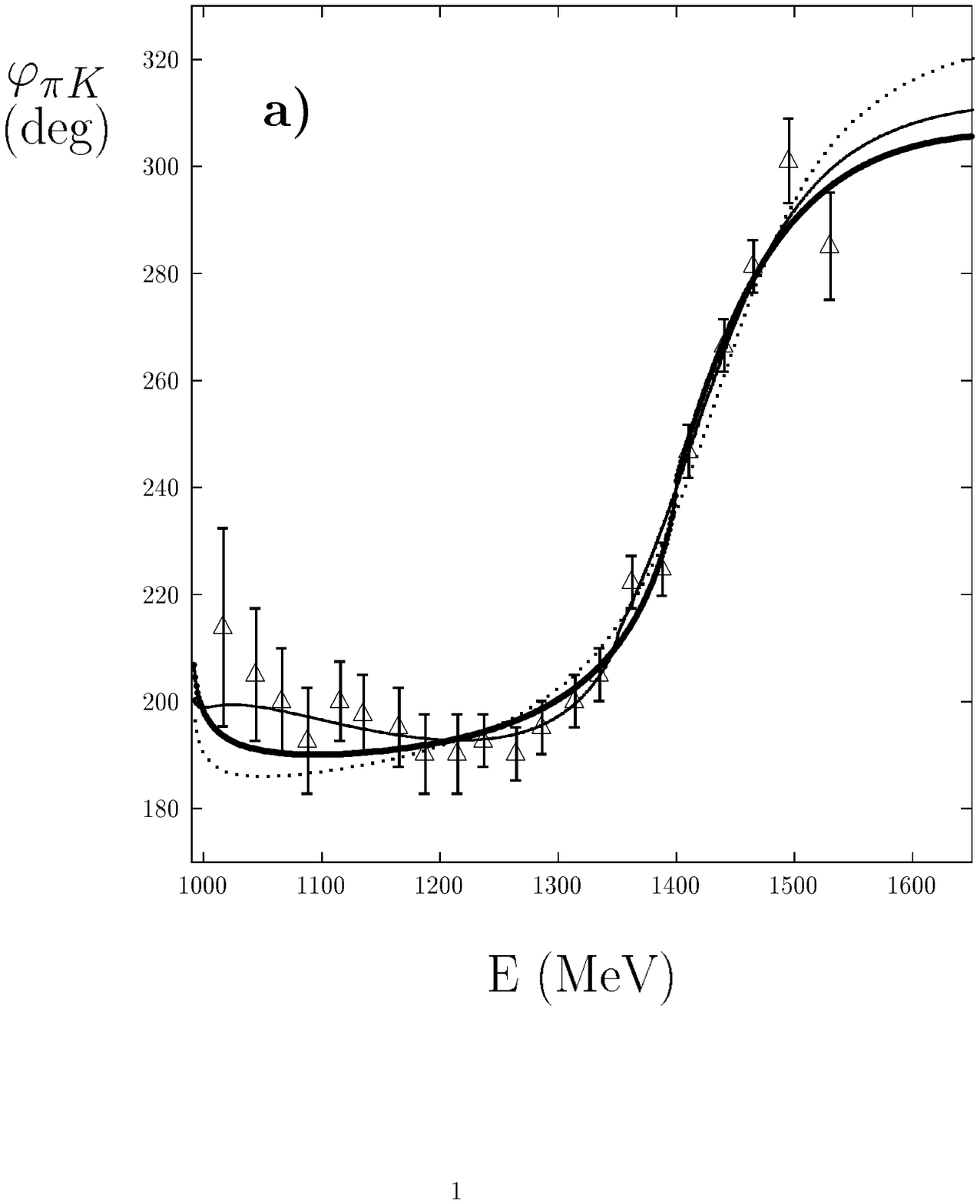}{10cm}{90}{220}{550}{710}{14cm} 
\xslide{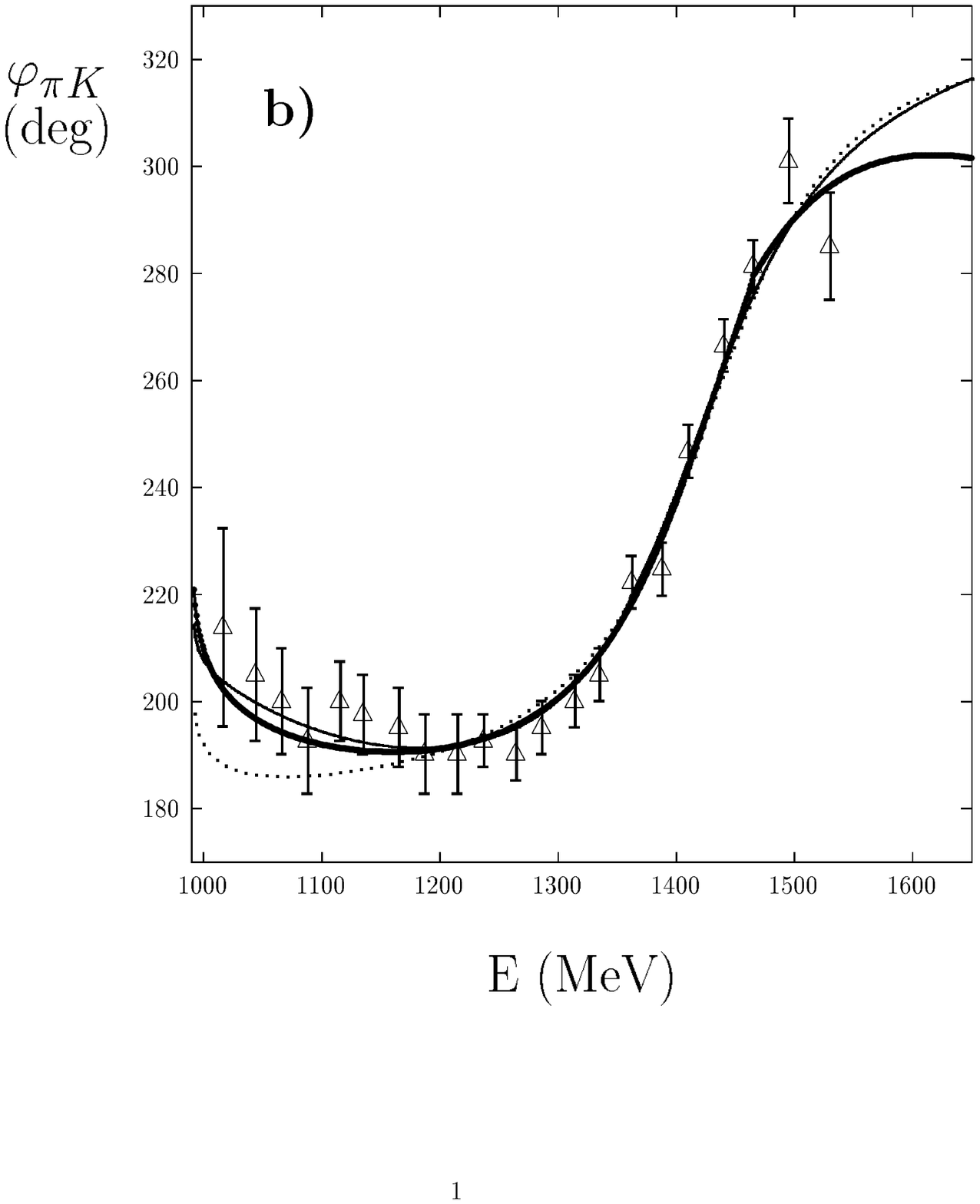}{10cm}{90}{220}{550}{710}{14cm} 
    \end{center}
  \end{figure}


\end{document}